\documentclass[aps,prb,nofootinbib,,eqsecnum,preprint,showpacs,amsmath,amssymb,floatfix,showpacs]{revtex4}  

\usepackage{graphicx}
\usepackage[utf8]{inputenc}
\usepackage{dcolumn}
\usepackage{bm}

\begin{document}
\title[]{Reverse Monte Carlo modeling in confined systems}

\author{V. Sánchez-Gil, E.G. Noya and E. Lomba}
\affiliation{Instituto de Qu{\'\i}mica F{\'\i}sica Rocasolano, CSIC, Serrano 119, E-28006 Madrid, Spain }
\date{\today}
\begin{abstract}
An extension of the well established Reverse Monte Carlo (RMC) method for
modeling systems under close confinement has been developed.
The method overcomes limitations induced by close confinement in systems
such as fluids adsorbed in microporous materials. As a test of the
method, we investigate a model system  of $^{36}$Ar adsorbed into
two zeolites with significantly different pore sizes: Silicalite-I (a pure silica form of ZSM-5 zeolite, characterized by
relatively narrow channels forming a 3D network)  at partial and full
loadings and siliceous Faujasite (which exhibits relatively wide
channels and large cavities). The model systems
are simulated using Grand Canonical Monte Carlo and, in each case, its structure
factor is used as input 
for the proposed method, which shows a rapid convergence and yields an
adsorbate microscopic structure in good agreement with that of the
model system, even to the level of three body correlations, when these
are induced by the confining media. The application to experimental systems is straightforward incorporating 
factors such as the experimental resolution and appropriate q-sampling, along the lines of previous 
experiences of RMC modeling of powder diffraction data including
Bragg and diffuse scattering. 
\end{abstract}

\pacs{61.05.-a, 68.43.Fg}
\maketitle
\section{Introduction}
Neutron and X-ray scattering techniques 
have been for years useful tools to gain a better
understanding of adsorption
processes\cite{Lang_1993_9_1852,Lang_1993_9_1846,Coulomb1994,Floquet2003,Floquet2007},
very specially in order to locate active sites and/or privileged
positions for  the adsorption of certain adsorbates. 
Given the small
ratio between adsorbate/adsorbent molecules, and since in many
instances the adsorbent exhibits a well defined crystalline structure,
one can expect a  diffraction pattern that will be dominated by
long range order features. This situation recalls the problem of 
modeling powder diffraction data to account for lattice and
magnetic disorder, which can be tackled by means of a Reverse Monte
Carlo (RMC) approach by 
direct calculation of the structure factor\cite{Mellergaard1999}. As
pointed out in Ref.\onlinecite{Mellergaard1999}, the well established
Rietveld refinement for modeling crystalline systems and its variants
mostly concentrate on the Bragg scattering whereas local disorder --which gives rise to diffuse scattering-- is
not considered. In the case of adsorption in crystalline microporous
materials, the adsorbate molecules do not necessarily exhibit
crystalline order. The Rietveld refinement can be applied using
hand-tuning to a certain degree when the number of adsorbate particles
per unit cell is relative low (see Refs.\onlinecite{Floquet2003} and
\onlinecite{Floquet2007} for examples of hydrocarbon adsorption in
Silicalite-I), and it is the approach of choice whenever the
adsorbate+adsorbent sample is fully crystalline, in which case the single
crystal method can be used (see
e.g. Refs.~\onlinecite{Koningsveld1987,Koningsveld1990,ACB_45_423,Kamiya2011}). This
approach would be certainly 
impractical when  there is a substantial degree of disorder. 

In this work, we are interested in the elucidation of adsorbate
structure in zeolites. These are materials with well defined microporous
geometry, in which corner-sharing AlO$_4$ and SiO$_4$ tetrahedra form
channels organized in 1D, 2D, and 3D networks accessible to different
adsorbate molecules. The crystalline structure of standard zeolites is
available from the literature\cite{Baerlocher2007}, and adsorbates
will induce changes in the diffraction spectra  due to either
 modifications in the
symmetry of the system or to the presence of disorder. From the discussion in the
preceding paragraph, it might seem  that the RMC approach of Mellegård 
and McGreevy \cite{Mellergaard1999}, as implemented in the RMCPOW
program\cite{Mellergaard2000} could be suitable to elucidate the
microscopic structure of adsorbates in the present instance. There are
however, a few aspects that suggest that a different approach is
needed. Firstly, in many cases, the changes induced in the zeolite
framework structure induced by the adsorbate are negligible (see
however Refs. \onlinecite{Koningsveld1987} and \onlinecite{ACB_45_423} as examples
in which relatively large adsorbates modify the spatial group of the
adsorbate). This implies that a substantial contribution to the
structure factor remains unchanged. On the other hand, if one tries to
blindly implement the standard Monte Carlo moves of ordinary RMC
approaches (basically molecular translations and/or rotations, or spin rotations to model magnetic disorder\cite{Mellergaard1998}) 
to molecules under tight confinement, most of the moves
will be rejected, by which the efficiency of the procedure will be
extremely poor as compared with that obtained in regular fluids and
glasses. The nature of our problem strongly suggests that the standard
translation/rotation moves must be complemented with 
particle creation/annihilation attempts that allow an efficient sampling.
It comes to our aid, that standard adsorption volumetry
experiments\cite{ASAP_greg_sing} provide with relative accuracy
estimates of the number of adsorbed molecules per unit cell of the
adsorbent. Bearing in mind this information, it is possible to
construct an efficient Reverse Monte Carlo procedure that can recover
the microscopic structure of the adsorbed fluid from powder
diffraction spectra and adsorption volumetry experiments. 

The aim of this work is to test the proposed approach, which we will
denote by N-Reverse Monte Carlo (N-RMC) method for several
model systems. The N-prefix underlies the fact that in this approach
the number of particles, N, is one of the variables to optimize. For
our testing purposes, we have generated the structure factor of
$^{36}$Ar adsorbed in two different zeolites, namely, Silicalite-I and siliceous Faujasite,
by means of Grand Canonical Monte Carlo (GCMC) simulations at different loadings.
Those systems have been studied experimentally by Llewellyn
and coworkers\cite{Lang_1993_9_1846,Maurin2005} and it is known that can 
reliably be modeled using GCMC simulations\cite{Maurin2005,Pellenq1995}. We will
see how the proposed N-RMC approach, with the sole input of the
relevant portion of the structure factor, the known zeolite structure, and an estimate of the number
of adsorbate molecules per unit cell can accurately render the microscopic structure
of the adsorbate in the course of a relatively short simulation run. 

The rest of the paper is sketched as follows. The essentials of the
method are introduced in Section \ref{method}. The most relevant
results are commented upon in Section \ref{results}. Conclusions and
future prospects are presented in Section \ref{conclusion}.

\section{Method}
\label{method}

\subsection{Implementation of the Reverse Monte Carlo method under confinement}
	
As mentioned before, information about the microscopic structure of the adsorbed fluid can be obtained from
neutron or X-ray powder diffraction measurements (see for example
Ref. \onlinecite{Lang_1993_9_1846}). In the case of neutron powder diffraction, we will be dealing with  an orientationally averaged
structure factor\cite{Mellergaard1999}:
\begin{equation}
\label{eq_sq}
S (q) = \frac{ 2 \pi^2}{N V<b>^2} \sum_{{\mathbf q'}} | F ({\mathbf q'}) |^2  
\delta (  q- q')/q'^2
\end{equation}
where  $N$ and $V$ are, respectively, the number of atoms and the volume of system 
(which in the case of a perfect crystal would reduce to those of the unit cell),
${\mathbf q'}$ are the allowed vectors in the reciprocal cell,
and $<b>$ is the average
of the coherent scattering lengths of the constituent atoms $b_j$. The $1/{q'}^2$ factor stems from the angular
integration over all the possible ${\bf q}'$ orientations in the
powder sample\cite{Mellergaard1999}. Finally, $F({\mathbf q})$ contains the correlations between the scattering nuclei and is given by:
\begin{equation}
F({\mathbf q}) = \sum_{j=1}^{N} b_j \exp(i{\mathbf q}{\mathbf R_j})
\end{equation}
where ${\mathbf R}_j$ denotes the position of the atom $j$ in the unit cell. When dealing with real 
experimental data , the $\delta$-function in Eq.~(\ref{eq_sq}) must be replaced by the instrument resolution function. As mentioned in Ref.~\onlinecite{Mellergaard1999} this can be any of the standard powder line shapes, e.g. a simple Gaussian distribution. 

In many cases of interest the
zeolite structure is hardly affected during the process of adsorption,
and for  practical purposes can be considered frozen. This is also a
very common approximation in simulation
studies\cite{ChemRev_2008_108_4125}. Along these lines,  in our
calculation the positions of the zeolite constituent atoms
will be kept frozen. 
Consequently, its contribution to the total
structure factor remains constant during the RMC simulation.
From an experimental point
of view, one typically measures the structure factor of the sample
with and without adsorbate. Since in our case, the zeolite structure
is well known, we will be working with the difference structure
factor,  
\begin{equation}
S_{diff}(q) = S_{total} (q) - S_{zeo-zeo}(q)
\label{sdiff}
\end{equation}
where $S_{zeo-zeo}$ is assumed to correspond to the empty sample. 
From Eq. (\ref{eq_sq})  the total structure
factor can be calculated in our case using the following expression:
\begin{equation}
S(q) = \frac{ 2 \pi^2}{N V<b>^2} \sum_{{\mathbf q'}} \frac{1}{q'^2}\left|
\sum_{j=1}^{N_{zeo}+N_{ad}} b_j \exp(i{\mathbf q'}{\mathbf R}_j)
\right|^2 \delta (  q-  q')
\end{equation}
where $N_{zeo}$ is the number of atoms of the zeolite and $N_{ad}$ is the
number of adsorbed atoms. 
It is easy to see that the three partial contributions to the total structure
factor can be calculated separately:
\begin{equation}
\label{szeo}
S_{zeo-zeo}(q) = \frac{ 2 \pi^2}{N V<b>^2} \sum_{{\mathbf q'}} \frac{1}{q'^2}\left|
\sum_{j=1}^{N_{zeo}} b_j \exp({i\mathbf q'}{\mathbf R}_j) \right|^2 \delta ( q-  q')
\end{equation}
\begin{equation}
\label{sad}
S_{ad-ad}(q) = \frac{ 2 \pi^2}{N V<b>^2} \sum_{{\mathbf q'}} \frac{1}{q'^2}\left|
\sum_{j=1}^{N_{ad}} b_j \exp({i\mathbf q'}{\mathbf R}_j) \right|^2 \delta (  q-  q')
\end{equation}
\begin{eqnarray}
\label{scros}
S_{zeo-ad}( q) &=& \frac{ 4 \pi^2}{N V<b>^2} \sum_{{\mathbf q'}} \frac{1}{q'^2} \left[ 
\left( \sum_{j=1}^{N_{ad}} b_j \cos({\mathbf q'}{\mathbf R}_j) \right)
\left( \sum_{j=1}^{N_{zeo}} b_j \cos({\mathbf q'}{\mathbf R}_j) \right) \right.\nonumber\\
&&+\left.
\left( \sum_{j=1}^{N_{ad}} b_j \sin({\mathbf q'}{\mathbf R}_j) \right)
\left( \sum_{j=1}^{N_{zeo}} b_j \sin({\mathbf q'}{\mathbf R}_j)  \right) \right] \delta (  q-  q')
\end{eqnarray}
As mentioned, we will only calculate the relevant contribution $S_{diff}$ --Eq.(\ref{sdiff})-- just adding Eqs.(\ref{sad}) and (\ref{scros}).
Note, however,  that in some cases the zeolite can 
undergo structural changes upon the adsorption of some molecules
(usually big aromatic molecules)
\cite{Koningsveld1990,Koningsveld1987,ACB_45_423}. Obviously, in those
cases the zeolite-zeolite contribution must be explicitly taken into account.

The core of the  RMC method reduces to performing random particle moves that are accepted or rejected depending on whether the newly generated trial structure of the fluid (measured in terms of the pair distribution function, $g(r)$, or the structure factor, $S(q)$) approaches a target  structure (usually an experimental $g(r)$ or $S(q)$). In the particular case that $S(q)$  is the reference property, the deviation from the target structure is measured using the statistical quantity,
\begin{equation}
\chi^2= \sum_{i=1}^{N_q} \frac{(S_{calc} (q_i) - S_{exp} (q_i))^2}{\sigma^2 (q_i)}
\end{equation}
where the sum runs over the $N_q$ discrete values of the wave vector $q_i$ for which the structure 
factor $S(q)$ is evaluated, and $\sigma (q_i)$ is the standard deviation of $S_{exp} (q_i)$, that takes into
account that experimental data carry different statistical uncertainties depending on the q-range of the measurements. 
In the standard RMC approach we will be dealing with translational or rotational trial movements. Following Ref.~\onlinecite{McGreevy1988}, 
the minimization of $\chi^2$ can be accomplished when the particle moves are accepted according to a probability given  by
\begin{equation}
\label{pacc}
P^{acc} = \min \left( 1, \exp\left(-\frac{\chi_{new}^2-\chi_{old}^2}{2}\right) \right)
\end{equation}
where $\chi_{new}$ and $\chi_{old}$ are the values of $\chi$ after and before the trial
move. In common with other optimization techniques such as simulating annealing and standard canonical Monte Carlo 
(that minimizes the system's internal energy), moves that worsen $\chi^2$  can also be accepted as long as they 
comply with the probability distribution  (\ref{pacc}). In this way,  the configurational space is adequately
sampled and chances for the procedure to get trapped in local minima are greatly reduced. 

Now, focusing on the problem of a system of tightly confined
particles, as is the case of adsorbates in zeolite channels, an
obvious problem with the scheme above described is the fact that most
translational moves (and rotations in the case of molecules) will be
rejected, due to overlaps with the zeolite framework. This means that,
even if  we are careful enough to generate an initial configuration of
non-overlapping adsorbate molecules, the very low diffusivity within
the channels and the high anisotropy of the medium, would render the
standard RMC method inefficient. 
Our approach to speed up the sampling consists on starting from the empty zeolite and, in addition to the usual
translational moves, also incorporate particle  insertion and
deletion trials.  The number of sample particles can be estimated from a variety of 
experimental sources, for example, from volumetric adsorption experiments,
and in standard RMC simulations is kept constant. In our approach,
the acceptance rule is modified so that
besides the minimization of $\chi^2$, a constraint on the 
target number of adsorbed particles, $N_{exp}$, is also included:
\begin{equation}
\label{pac2}
P^{acc} = \min \left( 1, \exp\left(-\frac{\chi_{new}^2-\chi_{old}^2}{2}- \frac{\Delta N_{new}^2-\Delta N_{old}^2}{2} \right)\right),
\end{equation}
where $\Delta N^2=(N-N_{exp})^2/\sigma_N^2$, $N$ being the instantaneous number of adsorbed particles and $\sigma_N$ 
the experimental uncertainty in $N_{exp}$. 

In this work the target $S(q)$ and $N_{exp}$ will be obtained from GCMC simulations
rather than from experiments and, therefore, both the target structure factor and the number of particles
are accurately known. However, we would like to explore the effect of their uncertainties on
the performance of the method. For that purpose we carried out simulations for several values of
$\sigma_N$ and $\sigma(q_i)=\sigma_S$, $\forall i$. For simplicity, we have chosen an uniform value for
the uncertainty of all q-values, but in real
experiments this is not necessarily so.
Note that the uncertainties play a similar role to the temperature in usual MC simulations,
i.e. $\sigma_S$ and $\sigma_N$ control
the equilibrium value and the magnitude of the allowed deviations of $\chi^2$ and $\Delta N^2$.
The lower the value of $\sigma_N$ the 
better the quality of the fit, but also the smaller the fluctuations allowed in $\chi^2$; and the same 
applies to the number of particles $N$ depending on the value of $\sigma_N$.

Due to the strong confinement effect imposed by the zeolite, insertion and deletion moves
are crucial to avoid the trapping of adsorbed atoms in particular regions of
the zeolite and therefore, will play a key role to sample efficiently the configurational space. 
Additionally, the performance of the RMC in confined media can be much improved by imposing
a bias in the insertion moves so that insertions are only attempted on those regions of
the zeolite accessible to the adsorbate \cite{JPC_93_97_13742}.  This
is sufficient in our case (a monoatomic adsorbate), but when dealing
with more  complex 
adsorbates, such as chain or aromatic molecules, more sophisticated
moves are needed. 
This is the same problem that one encounters in MC
simulations of complex molecules
in tightly confined media or at high densities.
It can be tackled by using configurational bias
moves\cite{FrenkelSmitbook} that have been designed to greatly enhance the
performance of sampling in the case of molecules with important steric
constraints (see Refs. \onlinecite{JPC_93_97_13742,Siepmann1992} for a comparison of the 
acceptance probability of purely random and various
types of biased displacement/insertion schemes in MC simulations).

Finally, as usual in the RMC method one has to define an exclusion core
around each of the sample particles. This core, that reflects the
effective size of the particle, is needed in order to avoid  unphysical
overlaps, either between the adsorbates or between the adsorbed atoms
and the zeolite framework. In our case, since we are dealing with model
Lennard-Jones particles this quantity can be defined rather easily. 
Similar to the usual RMC other constraints can be applied, for example, a
constraint on the adsorbate coordination number if many-body effects are known
to be important\cite{McGreevy2001}.  In the examples studied here,
many-body effects arise exclusively from the external field imposed
from the zeolite rather than from adsorbate-adsorbate interactions.
The target structure factor was obtained from MC simulations in which 
Ar-Ar interactions are simple pairwise Lennard-Jones potentials. In
the case of bulk systems interacting via pairwise potentials, it is known that the knowledge of the pair
distribution function determines uniquely the pair
potential\cite{Henderson1974}. Thus one should expect that in the
particular instance of pairwise interacting systems reproducing the
pair structure will guarantee an accurate representation of higher
order correlation functions without further constraints in the RMC
procedure. One must note however that in our case, effective many-body
effects on the Ar-Ar correlations are at play through the external
confining field. So the uniqueness of the structural resolution would
be in question, except for the fact that in the case of zeolites the
structure of the confining medium and its corresponding field are
accurately known. With this in mind, 
there will be no need to impose extra constraints on the procedure, as
it will be illustrated below. 

\subsection{Simulation details}

As mentioned, in order to assess the validity of the N-RMC
approach to study the structure of fluids under confinement,
we have considered as test cases the adsorption of argon in 
two zeolites with significantly different pore sizes, namely, Silicalite-I 
that is formed by a network of straight and sinusoidal pores of
diameter of about 5-5.5 {\AA}, and siliceous Faujasite that presents quite large cavities
with diameters of about 11.5 {\AA}. The first system was studied
experimentally\cite{Lang_1993_9_1846} by means of adsorption and
neutron scattering experiments. Nonetheless, for our test purposes, we
find more convenient to generate  the "experimental" structure factor
from a GCMC simulation. In this way, the target
structure is accurately known and we have access to all microscopic
structural quantities  of
relevance to compare with\cite{JCP_1991_94_3042}. Obviously this substantially simplifies the
problem, removing the experimental data treatment from the picture, or
the incorporation of the instrument resolution function (which should
replace the $\delta$-function in Eq.~(\ref{eq_sq})), and the
appropriate treatment of the discrete sampling of ${\bf
  q}$-space\cite{Mellergaard1999}. In our case we will be comparing
$S(q)$'s generated using identical simulation cells, by which all these
subtleties can be omitted. Obviously, this will not be the case when
dealing with real experimental data.  

\begin{figure}[!h]
\begin{center}
\includegraphics[width=160mm,angle=0]{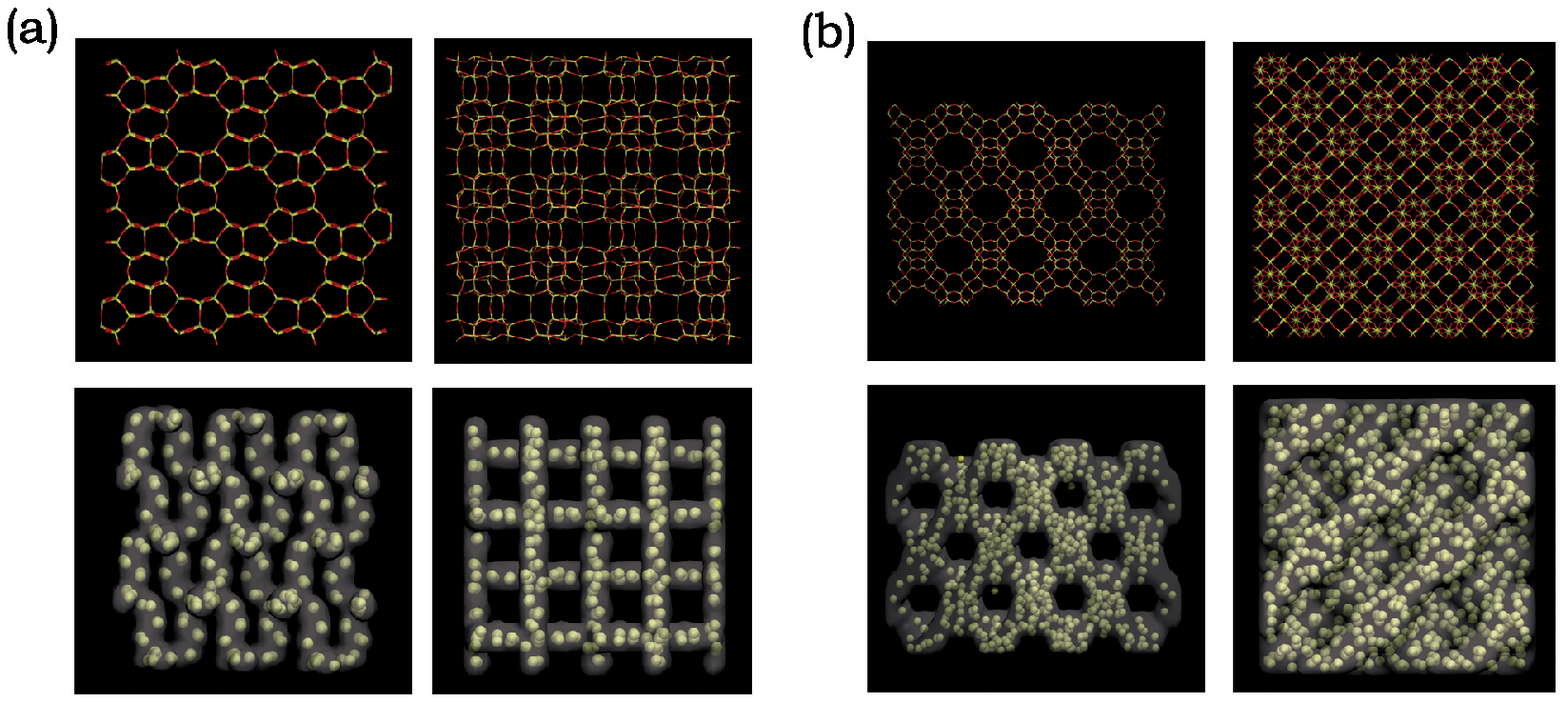}
\caption{\label{figure_structure}  Structure of the two zeolites considered in this work: (a) Silicalite-I and
(b) Faujasite. In each case, the two top panels show two different views of the zeolite structure and
the two bottom panels show the volume of the channels in a grey shadow and the adsorbed molecules
(at a loading of 32 atoms per unit cell in the case of Silicalite-I, and 100 atoms per unit cell in the
case of Faujasite).
For clarity purposes this figure shows smaller systems
that the one simulated in this work.}
\end{center}
\end{figure}

Explicitly, both in the GCMC and the N-RMC we have used a simulation box that contains
$4\times 4\times 6$ unit cells of the orthorhombic Silicalite-I\cite{Koningsveld1987} and 
$4\times 4\times 4$ for the Faujasite
(see the structures of these zeolites in Figure
\ref{figure_structure}).  
In the GCMC both the 
oxygen atoms in the zeolite and the argon atoms are modeled using
Lennard-Jones interactions. The parameters of the LJ model were chosen
from the bibliography\cite{Macedonia2000} and are given in Table \ref{tbl_parameters}. Silicon
atoms are surrounded by oxygen tetrahedra and therefore 
it is common not to assign a Lennard-Jones (LJ) center to them.
We used periodic boundary conditions and the LJ potential was truncated at a distance of 13 \AA.
GCMC simulations of argon adsorption were performed at 77K and in Silicalite-I at two different pressures
that lead respectively to loadings of about 25.5 and 32 argon atoms per unit cell, 
the latter corresponding to saturation. In Faujasite the study was
performed at a pressure corresponding 
to an intermediate loading of about 100 argon atoms per unit cell. 

\begin{table}[!h]
\centering
\caption{\label{tbl_parameters} Parameters of the Lennard-Jones model used
for the argon-argon and argon-zeolite interactions. }
\begin{tabular}{lcccc}
\hline\hline
	& & $\epsilon $/k$_B$ (K) & & $\sigma $ (\AA)  \\
\hline
Ar-Ar   & &        119.8       & &     3.405  \\      
Ar-O    & &        117.2       & &     3.121  \\      
 \hline\hline
\end{tabular} \\
\end{table}

	The simulated structure factor (subtracting the zeolite-zeolite contribution) 
averaged over a GCMC simulation of about 100,000 MC cycles was used as the target in the N-RMC run.  Here one
cycle is defined as $N_{ad}$ particle translations attempts plus one 
insertion and one removal attempt. We have defined the particle size
(or overlap distance) in the N-RMC as $\sigma_{\alpha\beta}(RMC) =
0.92\sigma_{\alpha\beta}$, (and $\sigma_{\alpha\beta}$ taken from
Table I) taking into account that the distance of minimum approach of
LJ particles is slightly less than the LJ $\sigma$ parameter. 

During the RMC simulation we
monitored the evolution of $\chi^2$ and
the number of particles. As it can be seen in Fig. \ref{fig_chi2}, at the beginning of the N-RMC run, the
quantity $\chi^2$ drops very rapidly whereas the number of atom increases,
until both quantities reach a plateau and  finally oscillate around 
an average value.  The magnitude of the oscillations in $\chi^2$ and $\Delta N^2$
can be controlled by the factors $\sigma_S$ and $\sigma_N$ that appear in the 
acceptance probability given in Eq.~(\ref{pac2}).

In addition to the straightforward comparison of the target
 and the RMC simulated $S(q)$'s, in our case a better insight
 of the method's performance can be gained by inspecting the partial
 distribution functions and the three body correlation functions. The partial distributions are
defined as: 
\begin{equation}
g_{{\alpha} {\beta}} (r) = \frac{n_{\alpha \beta} (r) }{\Delta V \rho_{\alpha}},
\end{equation}
where $n_{\alpha \beta}$ is the number of atoms of type $\beta $ at a distance
between $r$ and $r+\Delta r$ of a central atom of type $\alpha$, $\Delta V$ is 
the volume of a spherical shell between  $r$ and $r+\Delta r$, and $\rho_{\alpha}$ is the 
partial density of component $\alpha$. 
We have calculated the adsorbent-adsorbent (Ar-Ar) and the adsorbent-adsorbate (Ar-O) 
partial distribution functions. No particular information can be
extracted from the correlations involving Si atoms, since they all are  buried within the framework tetrahedra formed
by the oxygen atoms.
In order to investigate the three body
correlations we calculated the bond angle distribution, which is defined 
as the integral of the three body correlation function 
$g^{(3)}(r_1,r_2,\cos\theta)$ over the first coordination shell:
\begin{equation} 
\label{ftheta}
f (\theta) = 16 \pi^2 \int_0^{r_c} \int_0^{r_c} r_{13}^2 dr_{13} r_{23}^2 dr_{23} g(r_{12}) g^{(3)}(r_{13},r_{23},\cos\theta) ,
\end{equation} 
where we chose $r_c$ as the position of the first minimum of the 
pair distribution function. This function gives the distribution of angles
between pairs of nearest neighbors with respect to a central atom.
In this case we restricted our study to the bond angle distribution
for argon triplets.  From a practical point of view, this quantity
will be evaluated from the ensemble average of $\cos\theta_{132}$ histograms corresponding to
the 132 triplets of particles which fulfill $r_{13} < r_c$ and
$r_{23} < r_c$.

\section{Results}
\label{results}

\begin{figure}[!h]
\begin{center}
\includegraphics[width=140mm,angle=0]{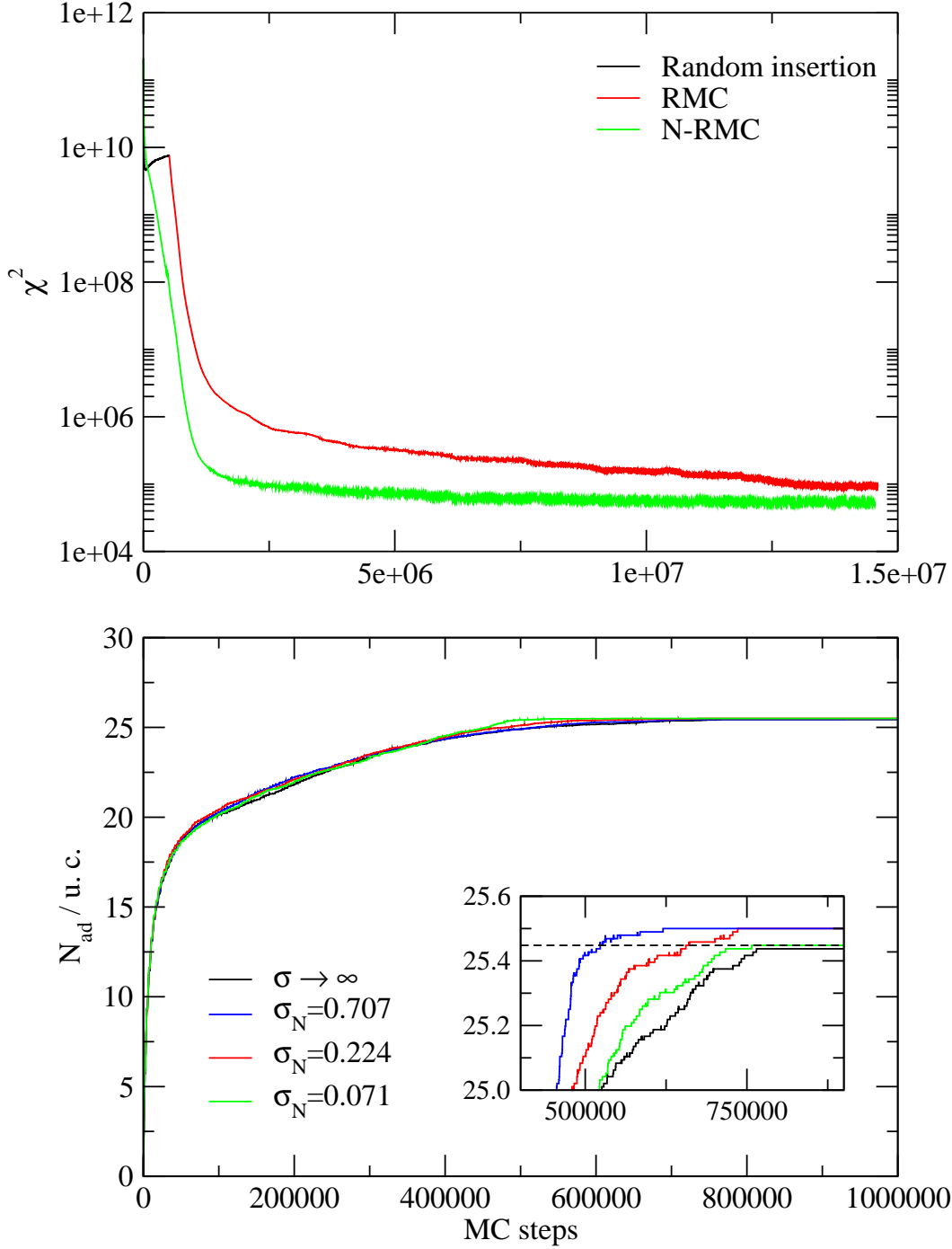}
\caption{\label{fig_chi2}  Evolution of $\chi^2$ and the 
number of adsorbed particles $N_{ad}$ per unit cell along the N-RMC simulation. 
The evolution of $\chi^2$ with the usual RMC code is also shown for comparison.
The black line shows the value of $\chi^2$ during the random insertion
of molecules used to generate the initial configuration for the usual
RMC algorithm. Along with the evolution of the number of particles for the
value of $\sigma_N$ used in this work (shown in green), the results for other values 
of $\sigma_N$ are also shown. The dashed black line in the inset shows the target $N_{exp}$.}
\end{center}
\end{figure}

We will start presenting the results for argon adsorbed in Silicalite-I.
For the case of a loading of about 25 molecules per unit cell and for the chosen
values of the uncertainties, $\sigma_S$ and $\sigma_N$, the N-RMC runs were fully converged
after 10$^7$ MC steps (see Fig. \ref{fig_chi2}). 
Initially the number of particles increased rapidly until it
reached the experimental value after about 6$\times 10^5$ MC steps. 
Beyond this point the number of particles
remains constant and particle insertion/deletion moves are no longer accepted.
A potential enhancement of the algorithm would be the implementation of coupled
insertion/deletion moves in which the former are guided by a cavity bias that takes
into account the location of adsorbate molecules within the zeolite accessible volume.
In the present instance this improvement has not been deemed necessary.

	For the particular case studied here and for the chosen value of
$\sigma_N$, the number of particles equilibrates exactly to $N_{exp}$ for $\sigma_N=\sqrt{0.005}\approx0.0707$.
For larger values of $\sigma_N$ the final number of particles is different (although not too far) 
from the experimental value. The fact that even when the constraint on the number of particles
is not included (which corresponds to the case of $\sigma_N\to \infty$) the final number of particles is
relatively close to the experimental one is related to the high
accuracy in $S(q)$. Note, however, that 
when using experimental data which are subject to larger uncertainties, the deviation from
the experimental number of particles will be quite large unless $\sigma_N$ is given a value consistent
with the experimental uncertainty. 
On the other hand, as mentioned before, the value of $\sigma_S$ was chosen 
as a compromise between the quality of the fit of $S(q)$ and an efficient sampling of
the configurational space. Here we used $\sigma_S\approx \sqrt{(V \langle b\rangle ^2)/(2\pi^2\times 2\times 10^5)}$.
The effect of the choice of $\sigma_S$ will be discussed in more 
detail below.

The evolution of $\chi^2$ in the usual RMC method (which only includes 
displacement attempts) is also shown for comparison in Fig. \ref{fig_chi2}.
When using the usual RMC algorithm one needs 
a procedure to generate an initial configuration with the experimental
number of adsorbed molecules, which can be obtained by random insertion of 
particles discarding those configurations that imply
adsorbate-adsorbate or adsorbate-zeolite overlaps. In this work this procedure was accelerated by trying
only insertions at positions of the zeolite accessible to the adsorbed particles.
As it can be seen in Fig. \ref{fig_chi2}, in this rather simple case that involves
spherical particles and a moderate density of the adsorbed fluid, an initial
configuration is obtained within about
5$\times10^5$ MC steps. For more complex molecules,
such as for example long chain alkanes, 
more sophisticated bias algorithms that enhance the probability
of insertion of particles will be needed to generate an initial configuration
in a reasonable amount of time\cite{JPC_93_97_13742,Siepmann1992}.
The RMC simulation started from this quasi-random configuration, which exhibits a quite large
value of $\chi^2$, seems to be converging to the same value as the N-RMC method
although at a much lower pace. Indeed, after $1.5\times10^7$ MC steps the RMC
method has not reached equilibrium yet, the average value of $\chi^2$ still 
decreasing.
The lower convergence of the RMC method
can be attributed to the low diffusion
of the particles in the zeolite. 
Note that the N-RMC method needs a slightly larger number of steps to reach the
experimental number of particles than the random insertion method.
However, the value of $\chi^2$ for the first
configuration with $N_{exp}$ molecules in the N-RMC,
although still quite high, is about two orders of magnitude lower than when particles
are inserted randomly, which indicates that particles are distributed already in 
a configuration much closer to the experimental one. The fact that virtually no 
particle exchange moves are accepted beyond this point indicates that indeed much of the
diffusion problems are overcome already in the filling process in the N-RMC.
It is quite remarkable that even for a simple system as that studied here
the N-RMC method speeds up the convergence considerably with respect to the
usual RMC method. As mentioned before, when dealing with complex molecules the
use of biased insertion/deletion moves is essential to sample efficiently
the phase space. In that instance the advantages of a N-RMC approach with
respect to a RMC method with simple translational/rotational moves (if the 
latter is feasible at all) should be more apparent.

\begin{figure}[!h]
\begin{center}
\includegraphics[width=140mm,angle=0]{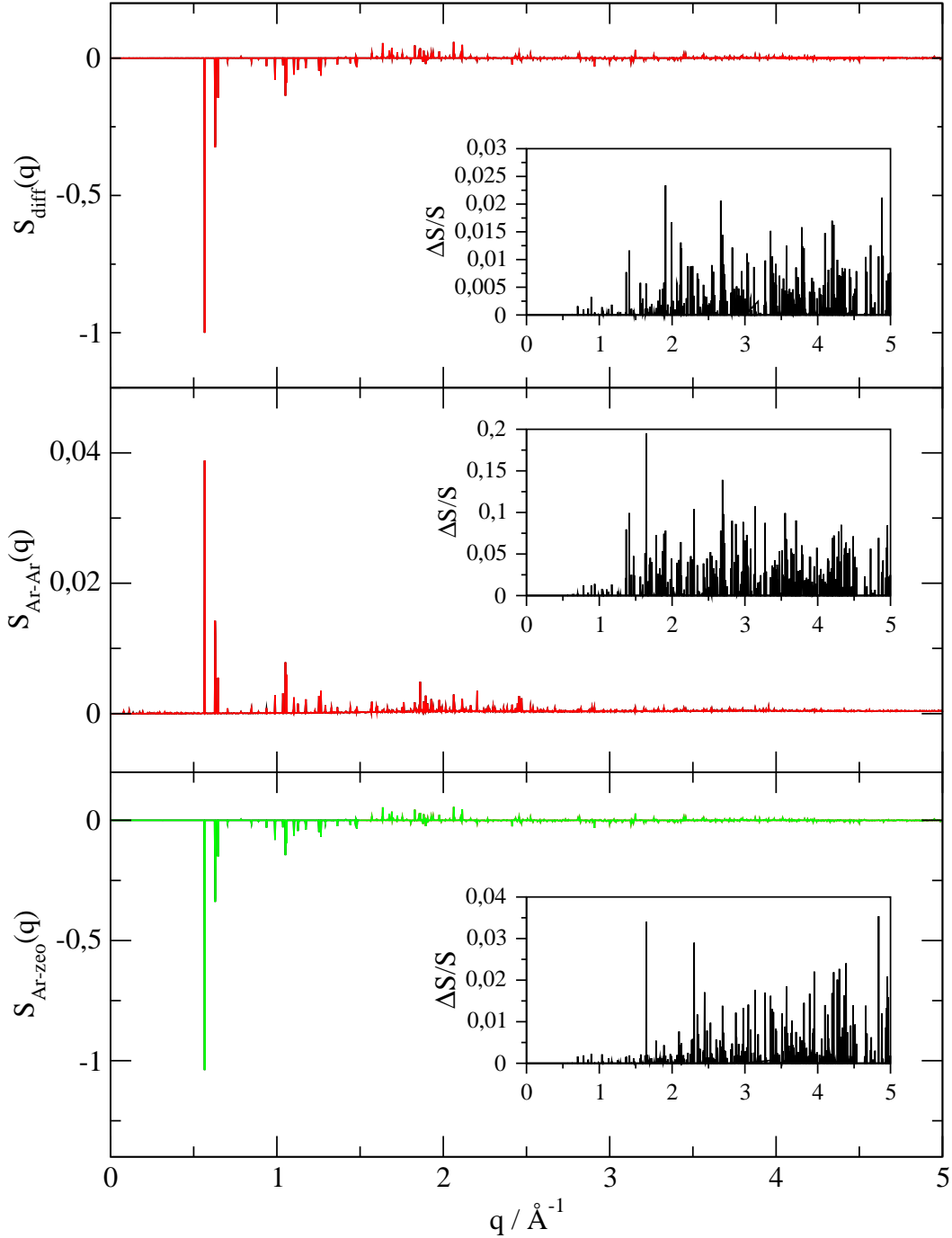}
\caption{\label{figure_structure_factor} Comparison of the target GCMC (black line)
structure factor and that obtained from the N-RMC (red line) in Silicalite-I at a loading
of about 25.5 $^{36}$Ar atoms per unit cell. Note that the zeolite-zeolite
partial structure factor has been subtracted so that only the sum of the 
argon-argon and argon-zeolite partial structure factors are used 
in the fit. The two separate partial structure factors are also shown.
The insets show the relative difference between the GCMC and N-RMC structure factors.}
\end{center}
\end{figure}

In Figure \ref{figure_structure_factor}, the
structure factor $S_{diff}(q)$ and its separate argon-argon  
and argon-zeolite contributions obtained from the N-RMC and the
target GCMC $S(q)$'s for a loading 
of 25 argon atoms per unit cell are shown. The low-energy neutron scattering
lengths have been taken from Ref.~\onlinecite{Lovesey1987}. Note that the
spiky appearance of both the target and fitted $S(q)$'s reflect the
finite number of $q$-vectors sampled and that no experimental
resolution function is taken into account. 
As mentioned before, all these factors must be explicitly incorporated
in order to fit experimental data\cite{Mellergaard1999}. Along
the N-RMC run, the number of particles rapidly converges to the
experimental value (see Fig. \ref{fig_chi2}), and the calculated and target $S(q)$'s are hardly
distinguishable, the relative difference between the GCMC and N-RMC
lying usually below 1\% (obviously for very low intensity peaks
relative errors can reach higher values,
but this corresponds to very small absolute errors). 

	Besides the good quality of the fit of  $S_{diff}(q)$, Figure \ref{figure_structure_factor}
shows that the same applies to the fit of the partial structure factor. Larger  relative differences
between the target and calculated argon-argon partial structure stem from the very low intensity of certain peaks of little relevance, and are also
an spurious result from the use of a constant $\sigma_S(q_i)$. 
The good quality of the calculated partial structure factors is an important result, since 
as both components (\ref{sad}) and (\ref{scros}) enter into $S_{diff}(q)$ some of their features could average out.
Therefore a good agreement in $S_{diff}(q)$ does not necessarily imply the same for its partial components.

\begin{figure}[!h]
\begin{center}
\includegraphics[width=140mm,angle=0]{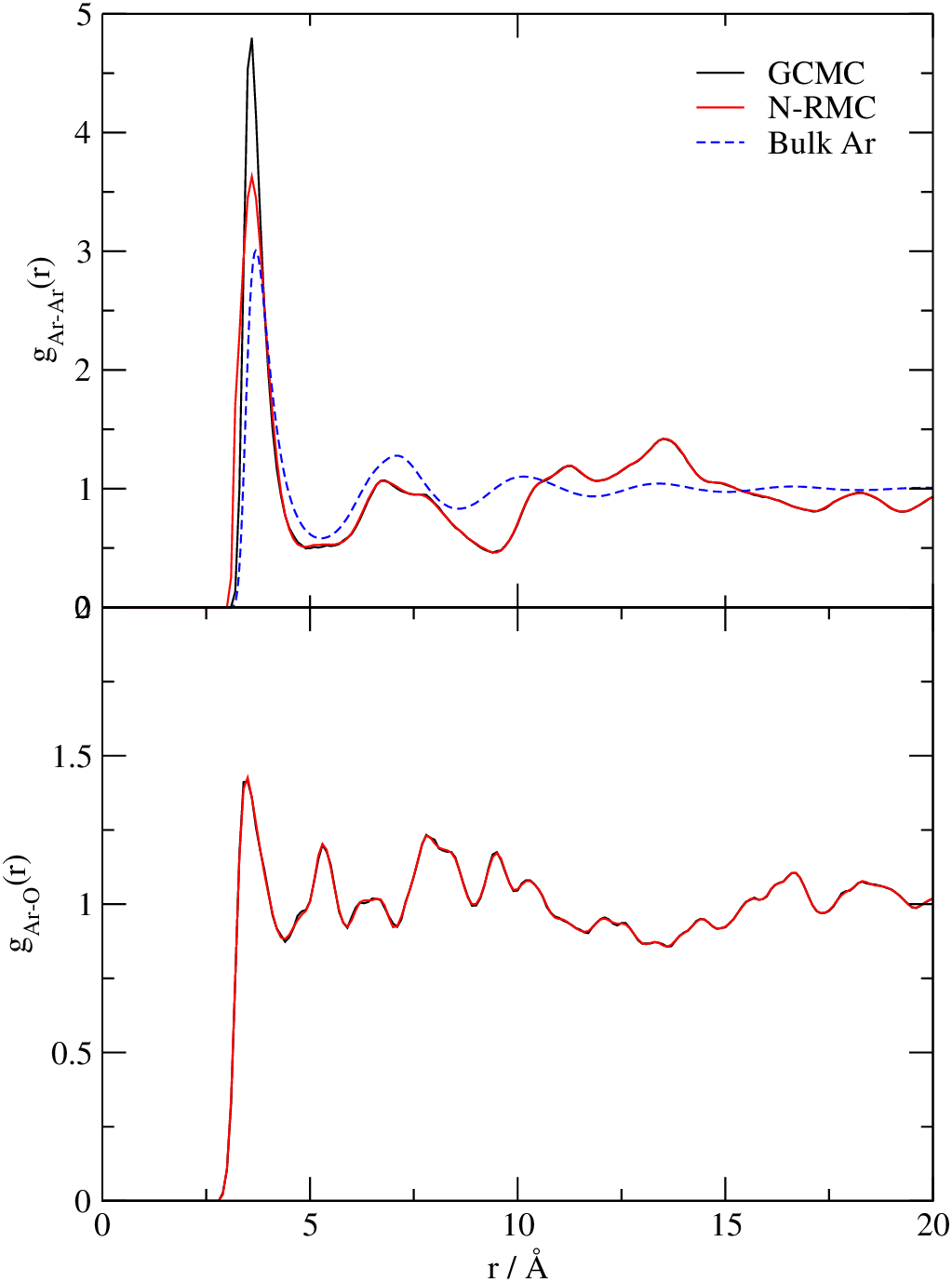}
\caption{\label{figure_rdf_25}  Comparison of the target (GCMC) and
the N-RMC partial Ar-Ar and Ar-O pair distribution functions in Silicalite-I at a
loading of 25.5 atoms per unit cell.}
\end{center}
\end{figure}

\begin{figure}[!h]
\begin{center}
\includegraphics[width=140mm,angle=0]{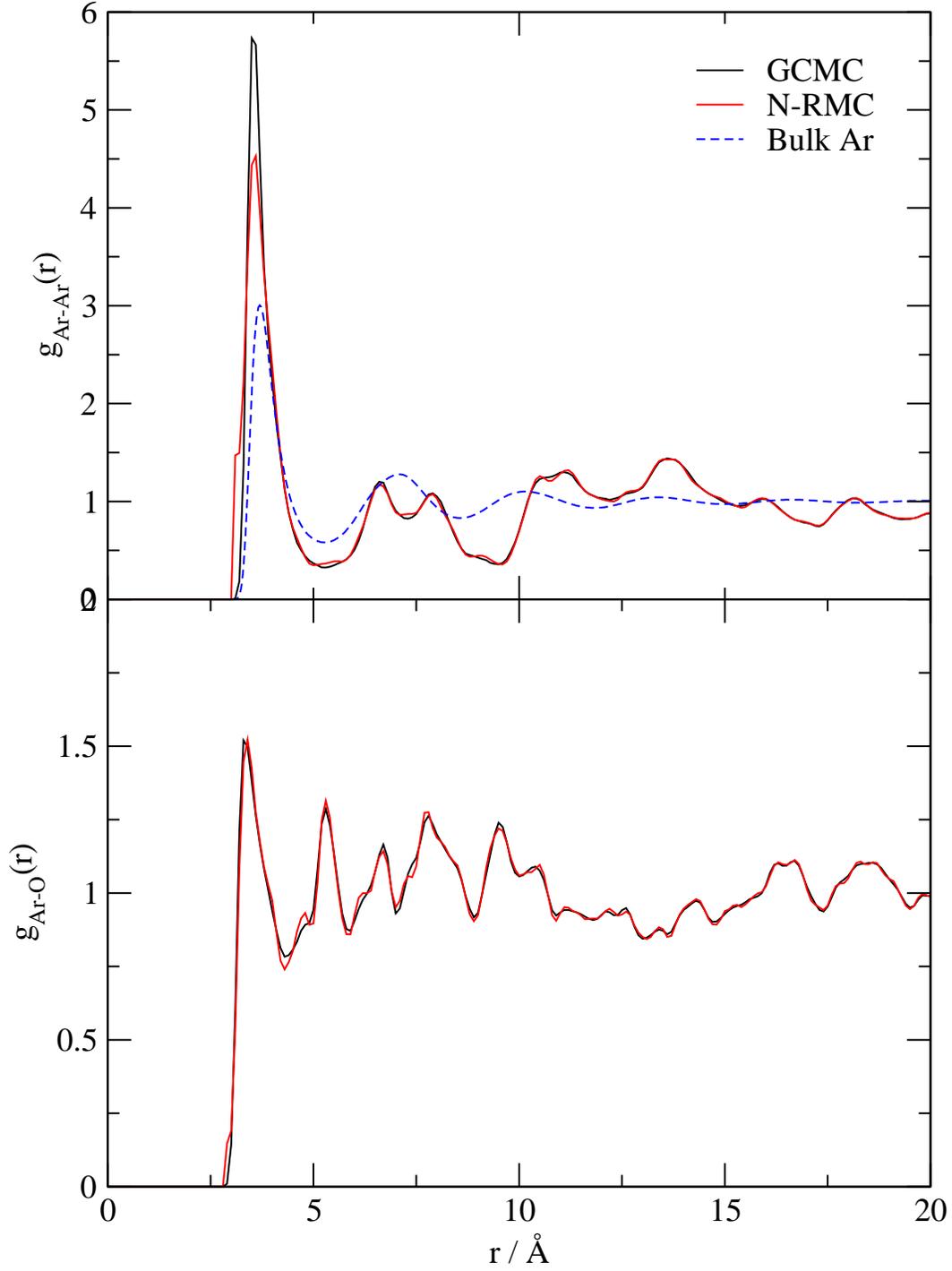}
\caption{\label{figure_rdf_32}  Comparison of the target (GCMC) and
the N-RMC partial Ar-Ar and Ar-O pair distribution functions in Silicalite-I at a
loading of 32 atoms per unit cell.}
\end{center}
\end{figure}

\begin{figure}[!h]
\begin{center}
\includegraphics[width=140mm,angle=0]{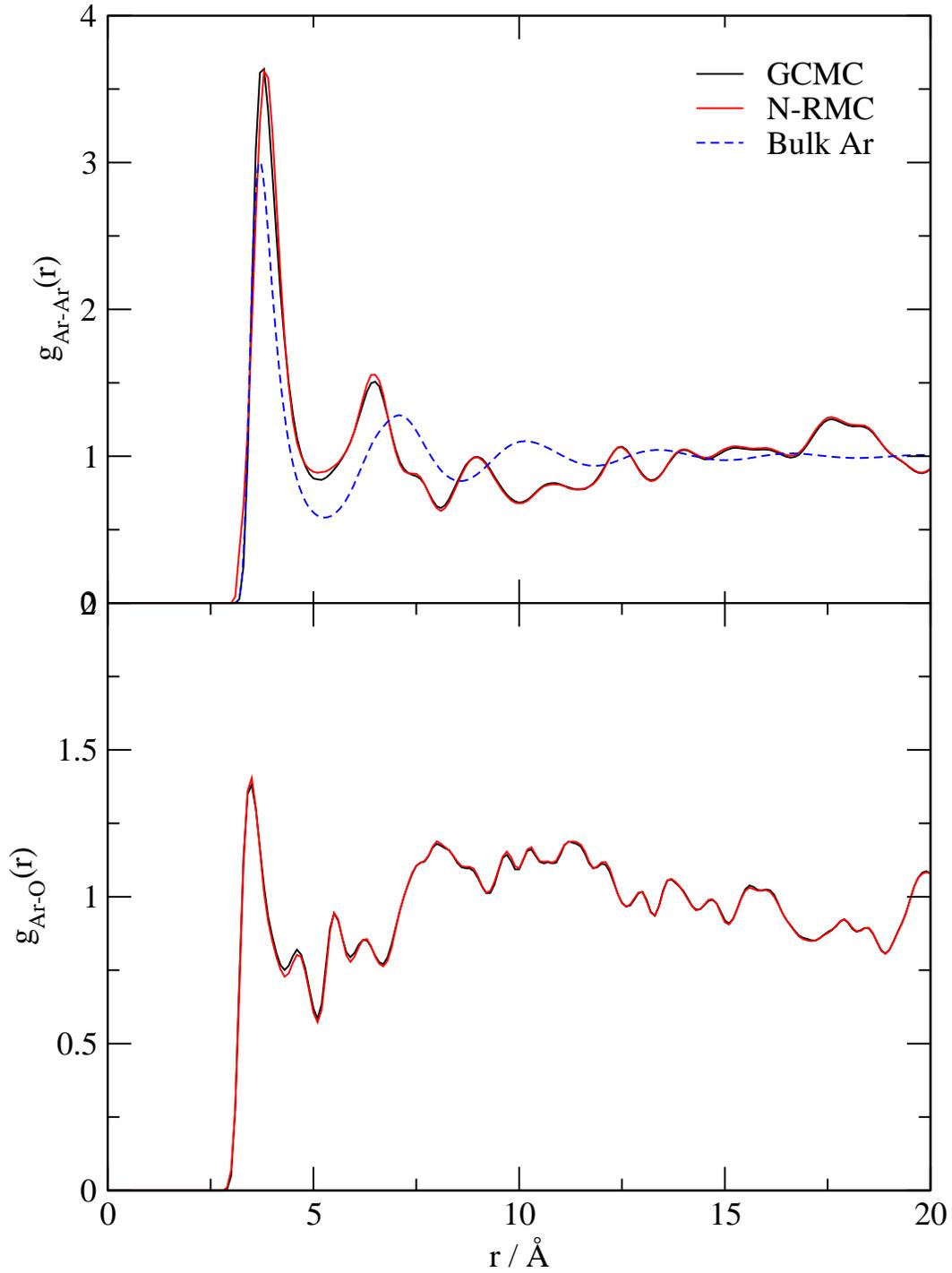}
\caption{\label{figure_rdf_fau}  Comparison of the target (GCMC) and
the N-RMC partial Ar-Ar and Ar-O pair distribution functions in Faujasite at a
loading of about 100 atoms per unit cell.}
\end{center}
\end{figure}

In order to get a better picture of the local order of the adsorbed fluid, and how 
this property is captured by the N-RMC approach, we analyze in Figures \ref{figure_rdf_25}, \ref{figure_rdf_32} and \ref{figure_rdf_fau}
the corresponding partial distribution functions extracted from both the N-RMC
and the GCMC simulations at the two adsorbate loadings in Silicalite-I (as an example of tight confinement) and Faujasite
(as an example of a more loose confinement). The bulk Ar distribution function evaluated
at the same temperature and at zero pressure is also shown for comparison.
For Silicalite-I, a first glance  at the  distribution functions shows  that 
the adsorbed fluid is very structured compared to the homogenous fluid, exhibiting order over quite long
distances, this order being induced by the topology of the zeolite channels. 
Note, however, that despite the long 
range of correlations observed in Figs. \ref{figure_rdf_25} and \ref{figure_rdf_32}, the sole intense 
peak corresponds to the nearest neighbor shell, this peak being higher and narrower than in bulk Ar, 
as is typically the case for fluids confined in narrow porous systems. The second peak on the other hand is split in two, the 
splitting being more apparent  for the higher loading.
The remaining peaks have a much lower intensity, though they extend over a wider range of 
distances than in the bulk fluid. This is in marked contrast with the situation observed when dealing with much larger adsorbate 
molecules at high loading (see Ref.\onlinecite{MMM_142_258}), in which the adsorbed molecules are forced to occupy 
highly correlated positions in the framework channels, giving rise to  much stronger interchannel adsorbate correlations. 
The argon-zeolite partial distribution functions are much less  structured, reflecting the small ratio of Ar vs. oxygen atoms.
In Faujasite, the larger size of the pores is reflected on a second fluid-like peak in the Ar-Ar distribution
function that occurs at shorter distances than the second peak in the bulk case (see Fig. \ref{figure_rdf_fau}).

When comparing the N-RMC and GCMC partial distribution functions, 
the overall good agreement for both argon-argon and argon-oxygen correlations 
is readily apparent, both in Silicalite-I (at the two loadings) and in Faujasite. 
The small differences on the first peak of Ar-Ar distribution function
arise due to the finite size of the simulation box. An accurate reproduction of the short-$r$ behavior
of $g(r)$ requires to have a detailed and accurate knowledge of the large $q$ 
behavior of $S(q)$. This in turn implies both the use of a large
system size that allows a finer 
sampling of $q$-space and the inclusion of a rather long  $q$-range in the fitting procedure.

\begin{figure}[!h]
\begin{center}
\includegraphics[width=140mm,angle=0]{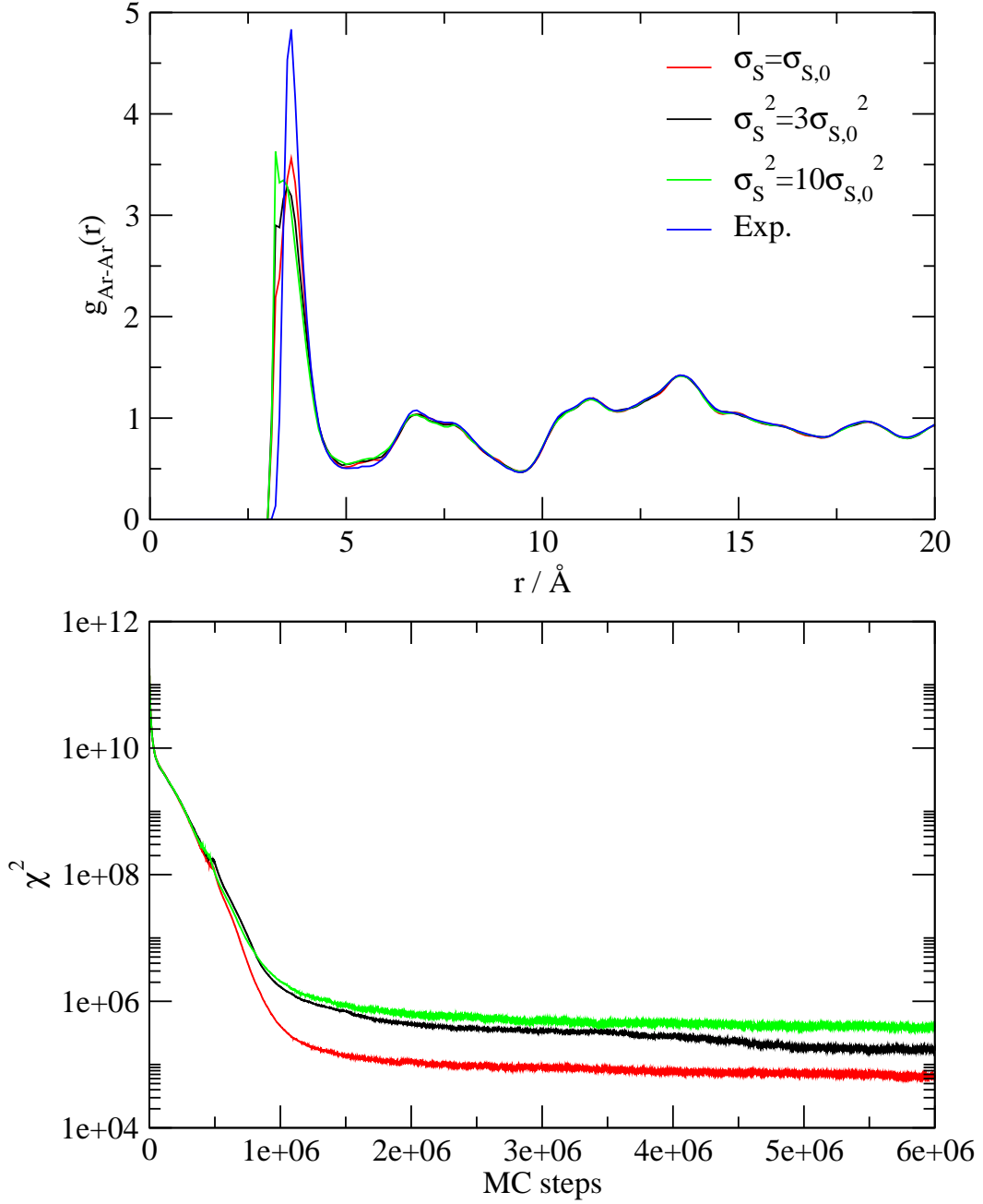}
\caption{\label{figure_temperature}  Effect of the $\sigma_S$ parameter on the resulting
pair distribution function (upper panel) and on the evolution of $\chi^2$ with
the MC step (lower panel) in the case of a loading of 25.5 atoms per unit cell
in Silicalite-I. $\sigma_{S,0}$ corresponds to the value used in this work,
$\sigma_{S,0} = \sqrt{(V \langle b\rangle^2)/(2\pi^2\times 2\times 10^5)}$.}
\end{center}
\end{figure}

Discrepancies in the first peak are more evident in Silicalite-I, i.e., in the system
with smaller pores that imposes a tighter confinement. This is not unexpected, a closer confinement
leads to a more solid-like behavior of the adsorbed fluid and relevant
features in the $S(q)$ extend to
larger $q$ than in systems with a more fluid-like behavior.
Obviously a small uncertainty in $S(q)$, $\sigma_S$ in Eq.~\ref{pac2}, 
is also required to accurately reproduce the first peak in the partial distribution functions.
As shown in Figure \ref{figure_temperature}, $\chi^2$ equilibrates to 
a lower value by decreasing $\sigma_S$, which means that the N-RMC $S(q)$ is closer to the target $S(q)$. 
As mentioned before, we chose  a value that allowed
us to sample the configurational space in a reasonable amount
of time and at the same time produces a fairly good quality fit.  We checked that the chosen value
corresponds to a very small relative error of about 0.001 \% for the most intense peak,
whereas the relative error grows up to 1\% for peaks with an intensity a thousand times smaller
than the most intense peak. Experimental data usually have a larger
statistical uncertainty of about a few percent (even in the more intense peaks). 
We checked the effect of a relative error on the target structure factor
and found that the distribution function was still reproduced with a very
good accuracy.
If any kind of medium-long range order builds up in the adsorbate within the
zeolite, N-RMC should be able to provide an appropriate microscopic picture
of it in consonance with the quality of the experimental data.

\begin{figure}[!h]
\begin{center}
\includegraphics[width=140mm,angle=0]{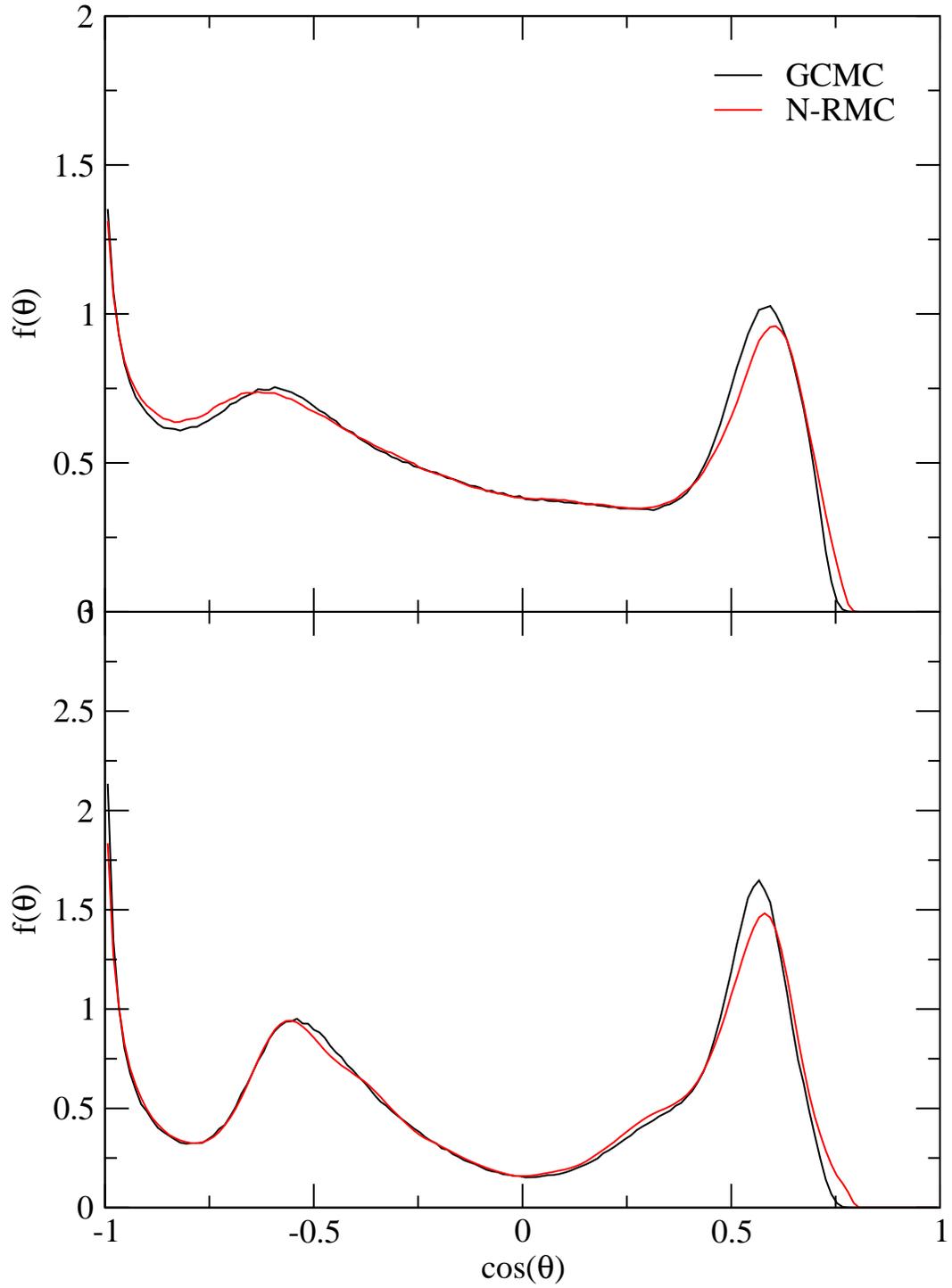}
\caption{\label{figure_bond_angle}  Comparison of the target (GCMC) and
the N-RMC bond angle distribution function --Eq.(\ref{ftheta})-- for argon triplets at a
loading of about 25.5 atoms per unit cell (top panel) and
32 atoms per unit cell (bottom panel) in Silicalite-I.}
\end{center}
\end{figure}

\begin{figure}[!h]
\begin{center}
\includegraphics[width=140mm,angle=0]{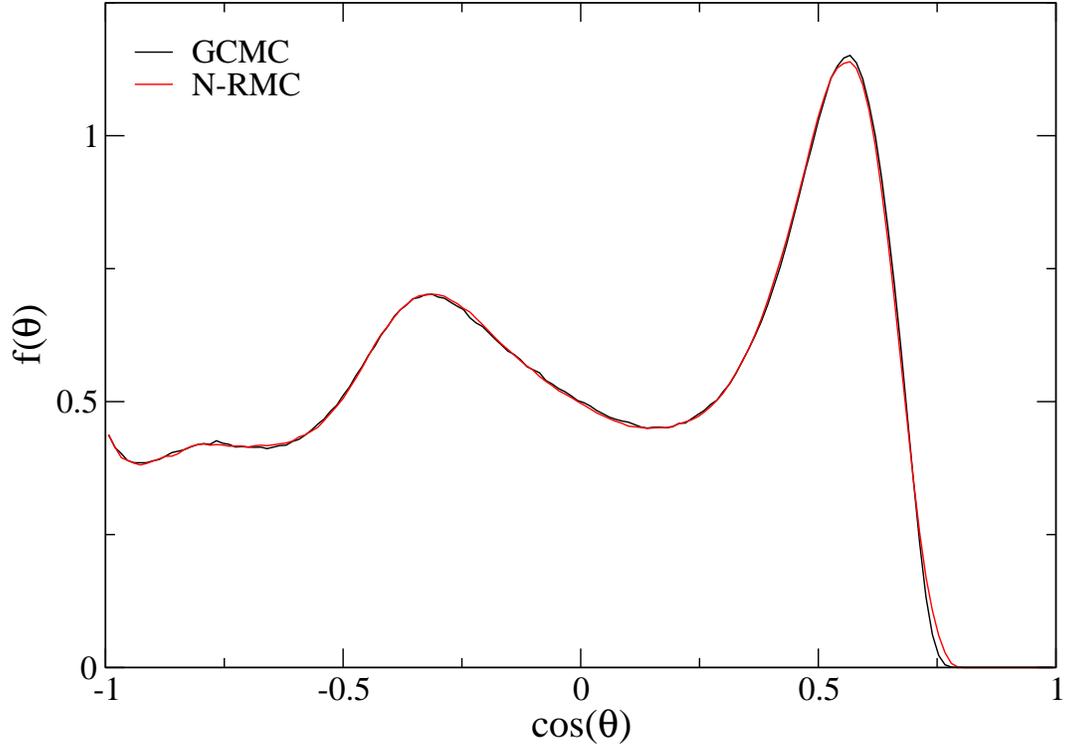}
\caption{\label{figure_bond_angle_fau}  Comparison of the target (GCMC) and
the N-RMC bond angle distribution function --Eq.(\ref{ftheta})-- for argon triplets at a
loading of about 100 atoms per unit cell in Faujasite.}
\end{center}
\end{figure}

Further insight into the structure of the adsorbed fluid can be gained
from the angular distribution function for triplets of argon atoms 
(see Figure \ref{figure_bond_angle} for Silicalite-I and Figure \ref{figure_bond_angle_fau} for Faujasite). 
As mentioned before, we integrated the triplet
correlation function up to the first minimum in the argon-argon partial distribution function ($\approx5$\AA).
For Silicalite-I the bond angle distribution is similar at
moderate and at high loadings, exhibiting peaks at roughly the same angles, but, as expected, 
the peaks are sharper at a high loading as a consequence of the higher density and reduced mobility
of the adsorbed atoms. The two peaks at high angles reflect the tendency
of the argon atoms to adopt local linear configurations imposed by the
confinement in the channels of the zeolite. 
The GCMC and N-RMC bond angle distribution functions agree very well at both loadings
except for some small discrepancies in the strong peak at short angles (these differences being connected to the
small error in the first peak of the Ar-Ar pair distribution function). 
On the other hand, the bond angle distribution
of Ar in Faujasite is more similar to the bulk LJ fluid, reflecting the larger pores in this zeolite.
In this case the agreement between the GCMC and the N-RMC bond angle distribution is almost perfect.
The correct prediction of the three body distribution function obtained here indicates that
the RMC method is able to capture the three-body correlations
induced mostly by the external field
created by the confining medium.  In
our particular case the structure of the confining medium is accurately known. Obviously, in those instances where the
intermolecular interactions of the adsorbate are strongly directional
with a significant influence of three body forces (e.g. in the case of
zeolite templated carbons\cite{Nishihara2009}) additional constraints
must be imposed along the RMC procedure, as it is customary in many
RMC applications (see for instance
Refs.~\onlinecite{JPCM_2005_17_3509,PRB_2008_77_195402} 
for particular applications to disordered carbon materials). In any case, as shown
for diatomic molecules and water, results obtained from RMC simulations need to be always
interpreted with caution, as the correct description of the pair
distribution function in real systems
does not necessarily mean a good reproduction of the higher order distribution functions\cite{JCP_1996_105_245,SM_1996_17_143}.
Nonetheless, as explained before, for tightly confined media, one would expect that geometric
effects play a more significant role. In those instances, the N-RMC
approach can be a very useful tool.

Before concluding we would like to comment on the range of $q$ used to fit
the $S(q)$. For the example presented here 
choosing a rather narrow range of $q$ ($q\leq5{\AA}$) was shown to be enough
to obtain a very good description of the adsorbed fluid. Indeed, simulations
including a broader range of $q$ did not lead to a substantial improvement of the
results. An appropriate sampling of the reciprocal space (by increasing
the simulation box more reciprocal vectors are included in the evaluation of S(q)) has
a higher impact on the quality of the results. The reason why including a higher range of $q$ has little influence
on the results is that oscillations
in the target $S(q)$ practically die out  for $q > 5{\AA}$ (see Fig. \ref{figure_structure_factor}). This can be understood
as the result of the relatively simple short range structure of the
adsorbed Ar, which must be recalled is one of the simplest fluids in
all respects. 
However, for systems in which the short range $g(r)$ displays
significant features (e.g. due to intramolecular correlations), fine long
range details of $S(q)$ cannot be neglected, and consequently
a broader range of $q$ must be included in the fitting procedure. When
dealing with real systems, these data can be obtained
from x-ray or neutron diffraction experiments, which currently allow to acquire fairly
high resolution data up to quite large values of $q$. 

\section{Conclusions}
\label{conclusion}
In summary, we have presented a simple extension of the Reverse Monte
Carlo method 
that enables the determination of the microscopic structure of fluids
under confinement. The success
of our test case study of a monoatomic fluid adsorbed into two well
known zeolites (Silicalite-I and Faujasite), evidences the performance of the
proposed method. Our approach can easily be extended to other systems,
even disordered porous materials, provided a previous study to
determine the structure of the adsorbent material is performed as
prerequisite. Complex molecular adsorbates can also be dealt with
resorting to bias sampling techniques. In a forthcoming publication we will
demonstrate the application of the method to determine the structure
of adsorbed aromatic hydrocarbons in various zeolites using both X-ray
diffraction data and volumetric and microcalorimetric adsorption
experiments as input for the N-RMC.

\section*{Acknowledgments}
The authors gratefully acknowledge the support from the 
Spanish MINECO (Ministry of Innovation and Economy) grant  
No. FIS2010-15502 and from the 
Autonomous Community of  Madrid under
Grant No. S2009/ESP/1691. The CSIC is also
acknowledged for providing support in the form of the project PIE (Proyecto Intramural Especial)
201080E120. V.S.G. also acknowledges
the CSIC for support of his work by means of a JAE program PhD
fellowship.


\section*{References}




\newpage

\end{document}